\documentstyle[fleqn,11pt,epsf]{article}

\catcode`\@=11
\long\def\@makefntext#1{
\protect\noindent \hbox to 3.2pt {\hskip-.9pt  
$^{{\eightrm\@thefnmark}}$\hfil}#1\hfill}		

\def\@makefnmark{\hbox to 0pt{$^{\@thefnmark}$\hss}}	
	
\def\ps@myheadings{\let\@mkboth\@gobbletwo
\def\@oddhead{\hbox{}
\rightmark\hfil\eightrm\thepage}   
\def\@oddfoot{}\def\@evenhead{\eightrm\thepage\hfil
\leftmark\hbox{}}\def\@evenfoot{}
\def\sectionmark##1{}\def\subsectionmark##1{}}

\newcounter{sectionc}\newcounter{subsectionc}\newcounter{subsubsectionc}
\renewcommand{\section}[1] {\vspace{12pt}\addtocounter{sectionc}{1} 
\setcounter{subsectionc}{0}\setcounter{subsubsectionc}{0}\noindent 
	{\tenbf\thesectionc. #1}\par\vspace{5pt}}
\renewcommand{\subsection}[1] {\vspace{12pt}\addtocounter{subsectionc}{1} 
	\setcounter{subsubsectionc}{0}\noindent 
	{\bf\thesectionc.\thesubsectionc. {\kern1pt \bfit #1}}\par\vspace{5pt}}
\renewcommand{\subsubsection}[1] {\vspace{12pt}\addtocounter{subsubsectionc}{1}
	\noindent{\tenrm\thesectionc.\thesubsectionc.\thesubsubsectionc.
	{\kern1pt \tenit #1}}\par\vspace{5pt}}
\newcommand{\nonumsection}[1] {\vspace{12pt}\noindent{\tenbf #1}
	\par\vspace{5pt}}

\newcommand{\textlineskip}{\baselineskip=13pt}

\def\abstracts#1#2#3{{
	\centering{\begin{minipage}{4.5in}\baselineskip=10pt\footnotesize
	\parindent=0pt #1\par 
	\parindent=15pt #2\par
	\parindent=15pt #3
	\end{minipage}}\par}} 

\renewenvironment{thebibliography}[1]
	{\frenchspacing
	 \ninerm\baselineskip=11pt
	 \begin{list}{\arabic{enumi}.}
        {\usecounter{enumi}\setlength{\parsep}{0pt}     
	 \setlength{\leftmargin 12.7pt}{\rightmargin 0pt} 
         \setlength{\itemsep}{0pt} \settowidth
	{\labelwidth}{#1.}\sloppy}}{\end{list}}

\newcounter{itemlistc}
\newcounter{romanlistc}
\newcounter{alphlistc}
\newcounter{arabiclistc}

\def\@citex[#1]#2{\if@filesw\immediate\write\@auxout
	{\string\citation{#2}}\fi
\def\@citea{}\@cite{\@for\@citeb:=#2\do
	{\@citea\def\@citea{,}\@ifundefined
	{b@\@citeb}{{\bf ?}\@warning
	{Citation `\@citeb' on page \thepage \space undefined}}
	{\csname b@\@citeb\endcsname}}}{#1}}

\newif\if@cghi
\def\cite{\@cghitrue\@ifnextchar [{\@tempswatrue
	\@citex}{\@tempswafalse\@citex[]}}
\def\citelow{\@cghifalse\@ifnextchar [{\@tempswatrue
	\@citex}{\@tempswafalse\@citex[]}}
\def\@cite#1#2{{$\null^{#1}$\if@tempswa\typeout
	{IJCGA warning: optional citation argument 
	ignored: `#2'} \fi}}

\def\@refcitex[#1]#2{\if@filesw\immediate\write\@auxout
	{\string\citation{#2}}\fi
\def\@citea{}\@refcite{\@for\@citeb:=#2\do
	{\@citea\def\@citea{, }\@ifundefined
	{b@\@citeb}{{\bf ?}\@warning
	{Citation `\@citeb' on page \thepage \space undefined}}
	\hbox{\csname b@\@citeb\endcsname}}}{#1}}

\def\@refcite#1#2{{#1\if@tempswa\typeout
        {IJCGA warning: optional citation argument
	ignored: `#2'} \fi}}

\def\refcite{\@ifnextchar[{\@tempswatrue
	\@refcitex}{\@tempswafalse\@refcitex[]}}

\def\pmb#1{\setbox0=\hbox{#1}
	\kern-.025em\copy0\kern-\wd0
	\kern.05em\copy0\kern-\wd0
	\kern-.025em\raise.0433em\box0}

\def\fnt#1#2{\footnotetext{\kern-.3em
	{$^{\mbox{\scriptsize #1}}$}{#2}}}

\font\tenrm=cmr10
\font\tenit=cmti10 
\font\tenbf=cmbx10
\font\bfit=cmbxti10 at 10pt
\font\ninerm=cmr9

\font\eightrm=cmr8

\def\qed{\hbox{${\vcenter{\vbox{			
   \hrule height 0.4pt\hbox{\vrule width 0.4pt height 6pt
   \kern5pt\vrule width 0.4pt}\hrule height 0.4pt}}}$}}

\begin{document}



\normalsize\textlineskip
\thispagestyle{empty}
\setcounter{page}{1}


\vspace*{0.88truein}

\centerline{III Workshop of DGFM/SMF, Nov. 28 -Dec. 3/99,
Le\'on, Gto., Mexico, [gr-qc/9912nnn]}
\vspace*{0.065truein}
\centerline{AN ERMAKOV STUDY OF $Q\neq 0$ EFRW MINISUPERSPACE OSCILLATORS}
\vspace*{0.035truein}
\vspace*{0.37truein}
\centerline{\footnotesize H. Rosu$^1$ and P. Espinoza$^{1,2}$}
\vspace*{0.015truein}
\centerline{\footnotesize\it 1. Instituto de F\'{\i}sica,
Universidad de Guanajuato, Apdo Postal E-143, Le\'on, Gto., Mexico}
\baselineskip=10pt
\centerline{\footnotesize \it 2. Centro Universitario de los Altos,
Universidad de Guadalajara, Lagos de Moreno, Jal., Mexico}
\vspace*{10pt}
\vspace*{0.225truein}

\vspace*{0.18truein}
\abstracts{
{\bf Abstract.} A previous work on the Ermakov approach for empty
FRW minisuperspace universes of Hartle-Hawking factor
ordering parameter $Q=0$ is extended to the $Q\neq 0$ cases.
}{}{}


\textlineskip                  
\vspace*{12pt}                 

\vspace*{1pt}\textlineskip	
\vspace*{-0.5pt}
\noindent


\noindent




\noindent





In a previous work, we presented the EFRW (empty FRW) minisuperspace oscillator
of $Q=0$  Hartle-Hawking factor ordering parameter in the Ermakov 
framework [\refcite{rer1}].
Here, we apply the Ermakov procedure to the same cosmological oscillators
for $Q\neq 0$.  


The EFRW Wheeler-DeWitt (WDW)
minisuperspace equation reads
\begin{equation} \label{1}
\frac{d^2\Psi}{d\Omega^2}+Q\frac{d\Psi}{d\Omega}-\kappa e^{-4\Omega}\Psi
(\Omega) =0~,
\end{equation}
where $Q$ will be considered as a free parameter,
the variable $\Omega$ is Misner's time, and $\kappa$ is the curvature
index of the FRW universe; $\kappa =1,0,-1$ for closed, flat, open universes,
respectively.
For $\kappa=\pm 1$ the general solution is expressed in terms of
Bessel functions, 
\begin{equation} \label{2}
\Psi _{\alpha}^{+}(\Omega) =
e^{-2\alpha\Omega}\left(C_1I_{\alpha}(\frac{1}{2}e^{-2\Omega})+
C_2K_{\alpha}(\frac{1}{2}e^{-2\Omega})\right)
\end{equation}
and
\begin{equation} \label{3}
\Psi _{\alpha}^{-}(\Omega) =
e^{-2\alpha\Omega}\left(C_1J_{\alpha}(\frac{1}{2}e^{-2\Omega})+
C_2Y_{\alpha}(\frac{1}{2}e^{-2\Omega})\right),
\end{equation}
respectively, where $\alpha =Q/4$.
The case $\kappa =0$ does not correspond to a parametric oscillator
and will not be dealt with here.
Eq.~(1) can be mapped in a known way 
to the canonical equations for a classical point
particle of 
mass $M=e^{Q\Omega}$, generalized coordinate $q=\Psi$,
momentum $p=
e^{Q\Omega}\dot{\Psi}$, (i.e., velocity $v=\dot{\Psi}$),
and identifying Misner's time $\Omega$ with the classical Hamiltonian time.
Thus, one is led to
\begin{eqnarray}
\dot{q}\equiv\frac{dq}{d\Omega}&=&e^{-Q\Omega}p~\\   
\dot{p}\equiv
\frac{dp}{d\Omega}&=&\kappa e^{(Q-4)\Omega}q~.
\end{eqnarray}
These equations describe the canonical
motion for a classical EFRW point universe as derived from the
`time'-dependent Hamiltonian
\begin{equation} \label{4}
H_{\rm cl}(\Omega)=
e^{-Q\Omega}\frac{p^2}{2}-\kappa e^{(Q-4)
\Omega}\frac{q^2}{2}~.
\end{equation}
The Ermakov invariant ${\cal I}(\Omega)$, which is algebraically built as
a constant of motion, can be written (for details see [\refcite{cs}])
\begin{equation} \label{6}
{\cal I}(\Omega)=
(\rho ^2)\cdot \frac{p^2}{2}-(e^{Q\Omega}\rho\dot{\rho})\cdot pq+
(e^{2Q\Omega}\dot{\rho} ^2 +\frac{1}{\rho ^2})\cdot\frac{q^2}{2}~, 
\end{equation}
where $\rho$ is the solution of the Milne-Pinney (MP) equation,
$
\ddot{\rho}+Q\dot{\rho}-\kappa e^{-4\Omega}\rho=\frac{e^{-2Q\Omega}}
{\rho ^3}$.
There is a well-defined prescription introduced by Pinney in 1950
of writing $\rho$ as a
function of the particular solutions of the corresponding parametric oscillator
problem, i.e., the modified Bessel functions in the EFRW case.
In terms of the function $\rho _{\pm}(\Omega)$ the EFRW Ermakov invariant reads 
\begin{equation} \label{10}
{\cal I} _{EFRW}^{\pm}=\frac{(\rho _{\pm} p-e^{Q\Omega}\dot{\rho} _{\pm}
q)^2}{2}
+\frac{q^2}{2\rho ^2_{\pm}}=
\frac{e^{2Q\Omega}}{2}\left(
\rho _{\pm} \dot{\Psi} _{\alpha}^{\pm}-\dot{\rho} _{\pm}
 \Psi _{\alpha}^{\pm}\right)^2
+\frac{1}{2}\left(\frac{\Psi _{\alpha}^{\pm}}{\rho _{\pm}}
\right)^{2}~.
\end{equation}
The calculation of ${\cal I} _{\rm EFRW}^{\pm}$ has been carried by using
linear combinations of Bessel functions [\refcite{pe}]
respecting the initial conditions 
for the MP function as given by Eliezer and Gray [\refcite{eg}].

As in our previous paper, 
we calculate the `time'-dependent generating function 
$S(q,{\cal I},\Omega)$ allowing one to
pass to new canonical variables for which ${\cal I}$ is chosen to be the
new ``momentum"
\begin{equation}   \nonumber         
S(q,{\cal I},\Omega)=e^{Q\Omega}\frac{q^2}{2}\frac{\dot{\rho}}{\rho}+
{\cal I}{\rm arcsin}\Bigg[\frac{q}{\sqrt{2{\cal I}\rho ^2}}\Bigg]
         +\frac{q\sqrt{2{\cal I}\rho ^2-q^2}}{2\rho ^2}~, 
\end{equation}
where we have put to zero the constant of integration.
The canonical variables
are now $q_1=\rho \sqrt{2{\cal I}}\sin \theta$
and
$p_1=\frac{\sqrt{2{\cal I}}}{\rho}\Big(\cos \theta+
e^{Q\Omega}\dot{\rho}\rho\sin \theta\Big)$.
The dynamical angle will be
$\Delta \theta ^{d}=
\int _{\Omega _{0}}^{\Omega}
\langle\frac{\partial H_{\rm{new}}}{\partial {\cal I}}\rangle
d\Omega ^{'}=
\int _{0}^{\Omega}[\frac{e^{-Q\Omega '}}{\rho ^2}-
\frac{1}{2}\frac{d}{d\Omega ^{'}}
(e^{Q\Omega ^{'}}\dot{\rho}\rho)+e^{Q\Omega ^{'}}
\dot{\rho}^2]d\Omega ^{'}$,
whereas the geometrical angle reads
$\Delta \theta ^{g}=\frac{1}{2}\int _{\Omega _0}^{\Omega}
[\frac{d}{d\Omega ^{'}}
(e^{Q\Omega ^{'}}\dot{\rho}\rho)-2e^{Q\Omega ^{'}}
\dot{\rho}^2]
d\Omega ^{'}$.
Thus, the total change of angle is
$\Delta \theta =\int _{\Omega _{0}}^{\Omega}\frac{e^{-Q\Omega ^{'}}}
{\rho ^2}
d\Omega ^{'}$.
On the Misner time axis, going to $-\infty$ means going to the origin of the
universe, whereas $\Omega _{0}=0$ means the present epoch. Using
these cosmological limits one can see that the total change
of angle $\Delta \theta$ during the cosmological evolution in $\Omega$ time
is up to a sign the Laplace transform of parameter $Q$
of the inverse square of the MP function, $\Delta \theta=
-L_{1/\rho ^{2}}(Q)$.

We end up with the possible interpretation of the Ermakov 
invariant in empty minisuperspace cosmology. If one makes an adiabatic series 
expansion of the invariant, the leading term that defines the adiabatic regime
gives the number of created `quanta' and there were authors giving classical
descriptions of the cosmological particle production in such terms
[\refcite{pp}]. On the 
other hand, a simple physical interpretation of the Ermakov invariant has been
introduced by Eliezer and Gray, who associated it with the angular momentum of 
the so-called 2D auxiliary motion [\refcite{eg}]. Thus, we claim that in 
EFRW minisuperspace cosmologies the classical Ermakov invariant gives the 
number of auxiliary angular momentum adiabatic excitations 
with which the universe is created at the initial singularity.

In conclusion, we extended here our previous 
Ermakov procedure for the EFRW WDW parametric equation from $Q=0$ 
to the $Q\neq 0$ cases.
Plots of the relevant quantities are displayed in the following.

\bigskip

\bigskip


\bigskip

\vskip 1ex
\centerline{
\epsfxsize=190pt
\epsfbox{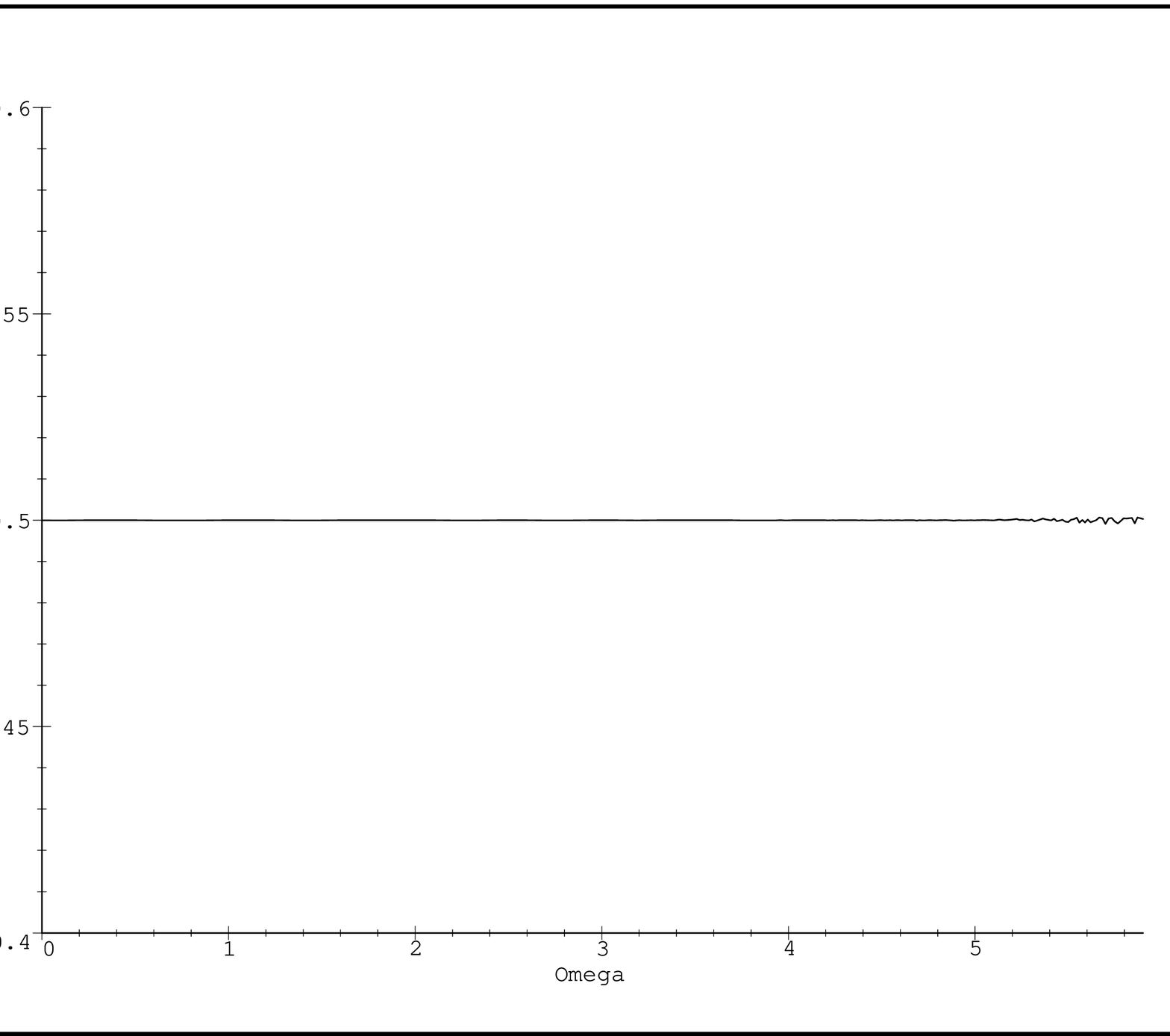}}
\vskip 2ex

\bigskip

\centerline{\footnotesize Fig.~1: ${\cal I} _{EFRW}^{+}(\Omega)$  
for the $Q=3$ case, corresponding to a singularity with
one auxiliary angular momentum excitation. }

\vskip 1ex
\centerline{
\epsfxsize=190pt
\epsfbox{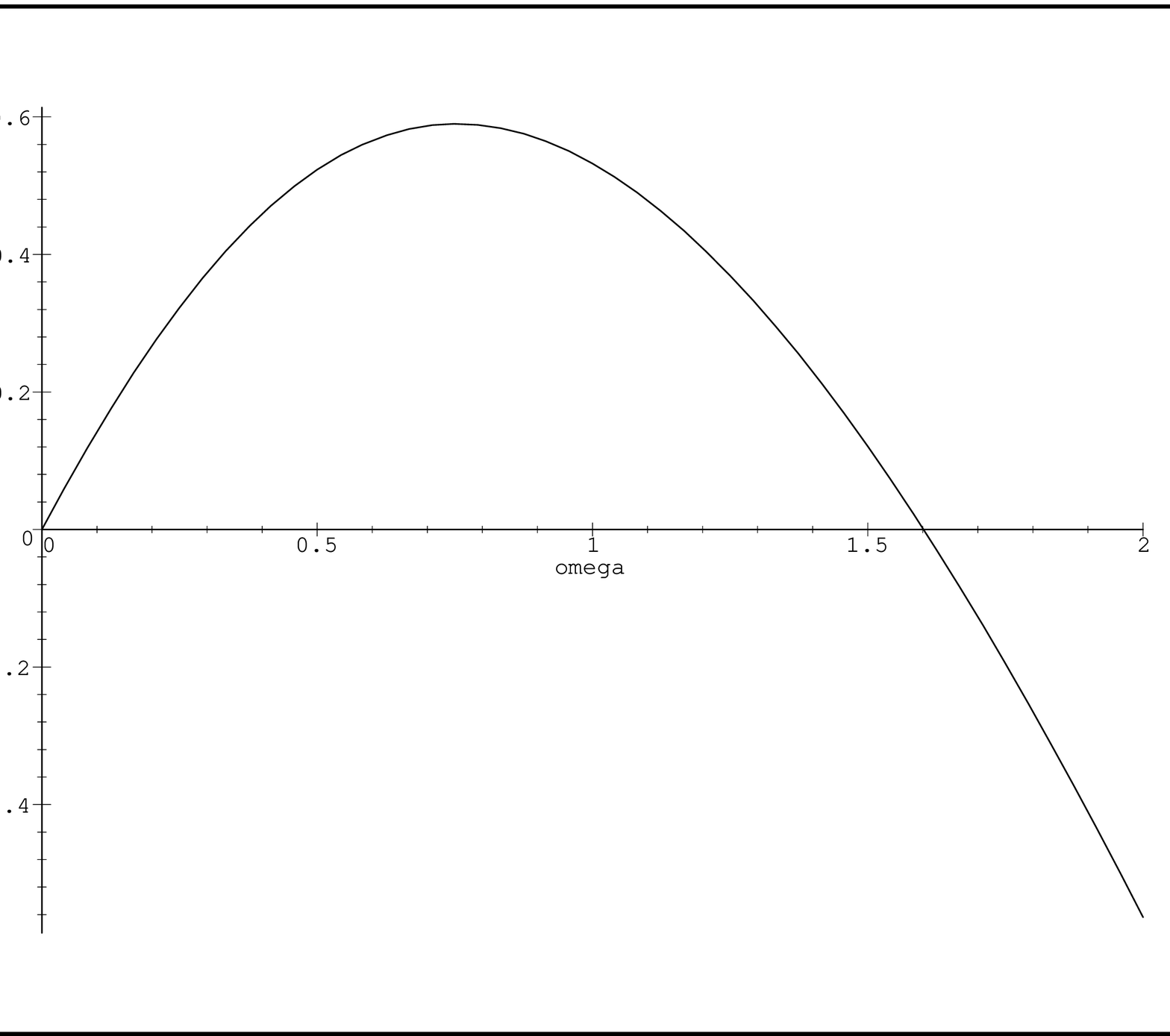}}
\vskip 2ex

\bigskip

\centerline{\footnotesize Fig.~2: The dynamical angle as a function of $\Omega$
for the closed EFRW model and $Q=1$.} 

\bigskip

\vskip 1ex
\centerline{
\epsfxsize=190pt
\epsfbox{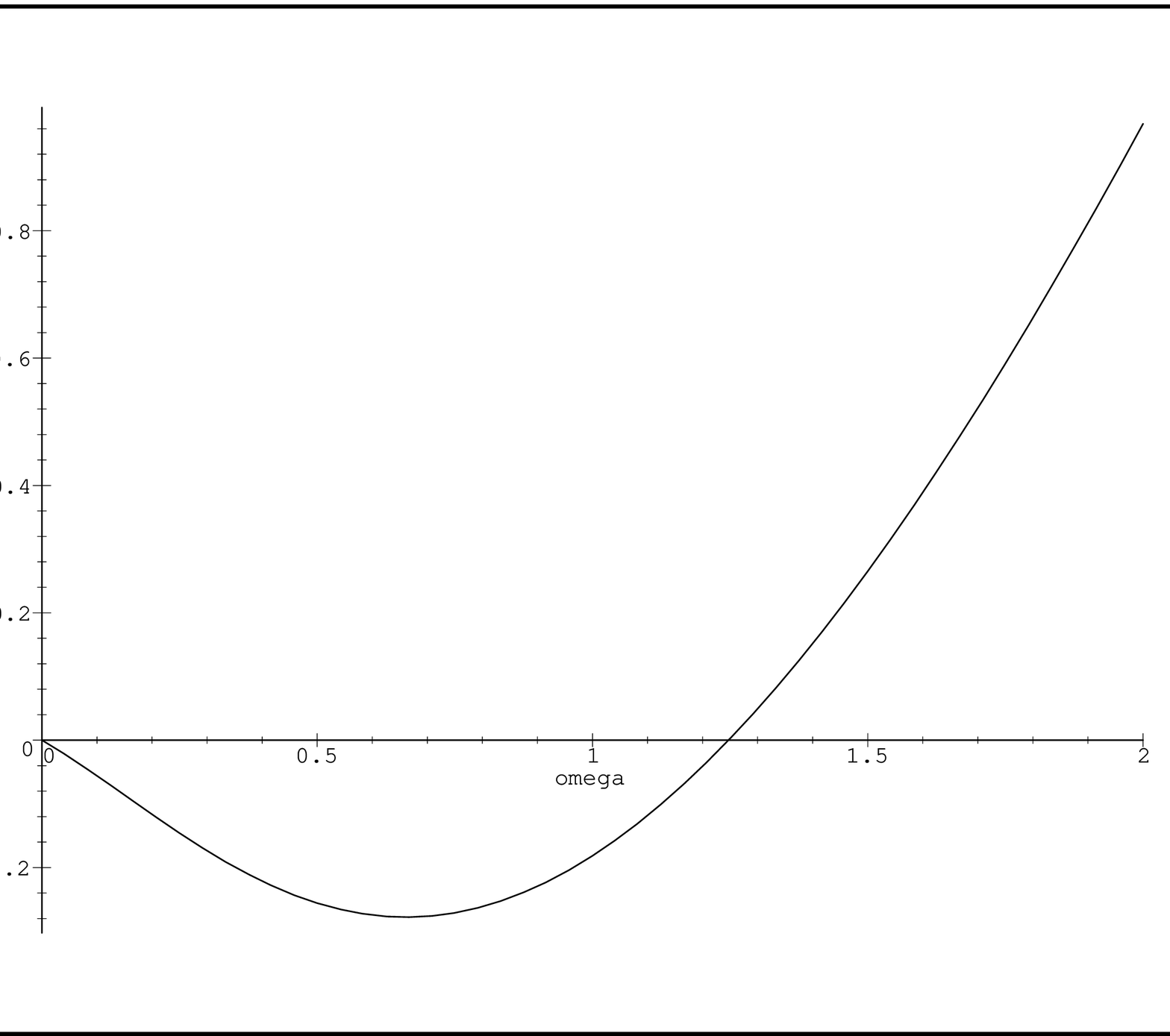}}
\vskip 1ex


\centerline{\footnotesize Fig.~3: The geometrical angle for the same case.} 

\bigskip

\vskip 1ex
\centerline{
\epsfxsize=190pt
\epsfbox{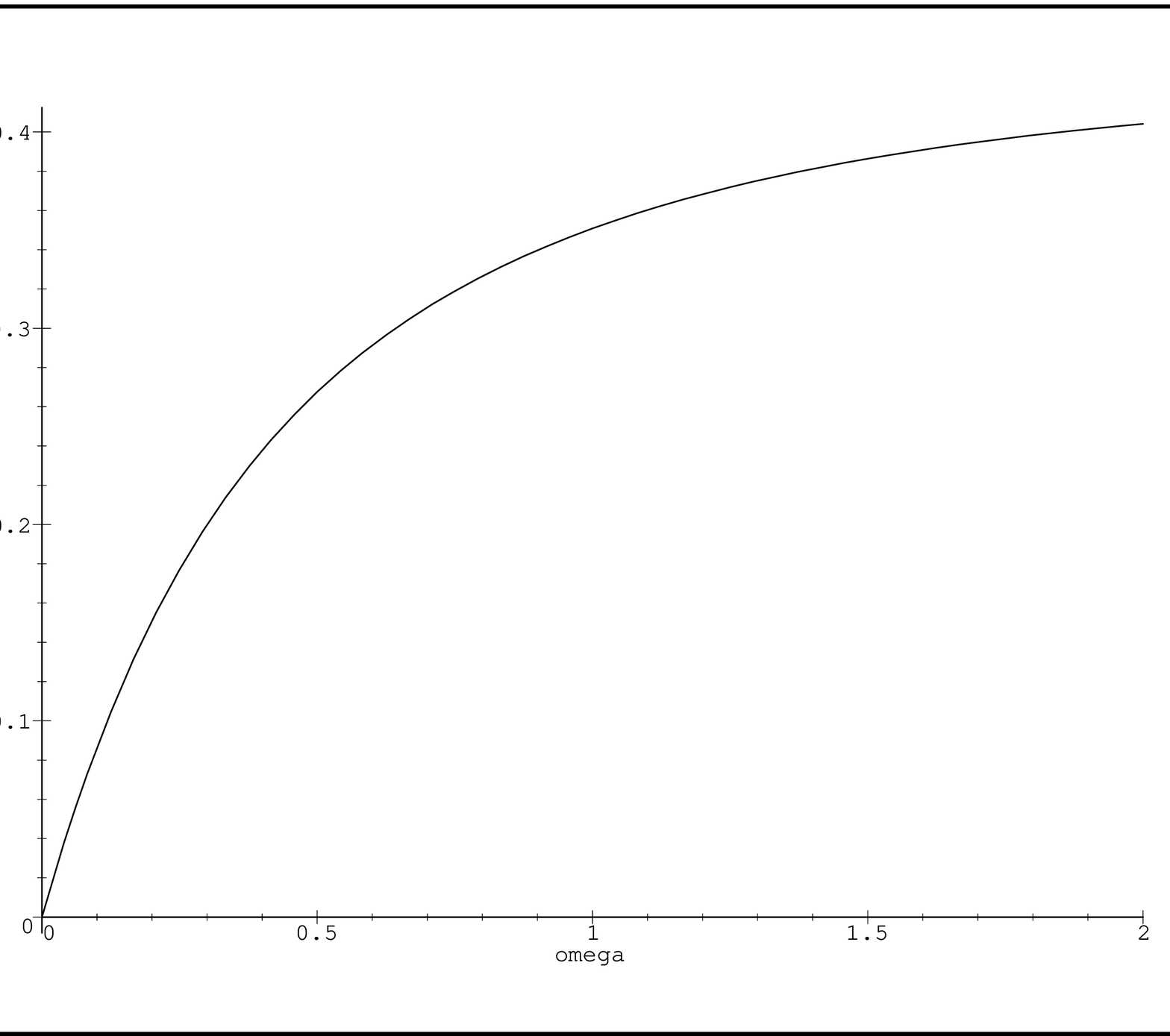}}
\vskip 1ex


\centerline{\footnotesize Fig.~4: The total angle as a function 
of $\Omega$ for the same case.}


\bigskip


\bigskip

\bigskip

\noindent
This work was partially supported by the CONACyT Project 458100-5-25844E.


\nonumsection{References}

\end{document}